# Machine Learning Approaches to Point Defects in Non-Metallic Materials: A Review of Methods

Yu Kumagai[1,2,*] and Shin Kiyohara[1]

[1] *Institute for Materials Research, Tohoku University, Sendai, Japan*

[2] *Organization for Advanced Studies, Tohoku University, Sendai, Japan*

*Corresponding author: yukumagai@tohoku.ac.jp

**We review recent machine-learning (ML) approaches for point defects in non-metallic materials, with an emphasis on defect formation energies. Existing studies largely fall into two categories: direct ML models that predict defect energetics from local structural representations, and machine-learning potentials (MLPs) that approximate the defect-containing potential energy surface. We summarize key achievements as well as persistent bottlenecks, emphasizing that dataset quality often dominates practical model performance. We further identify charged-defect formation energies as a central frontier, where Fermi-level alignment, finite-size corrections, and long-range electrostatics must be handled carefully and consistently to enable meaningful comparisons and transferable predictions across different materials.**

## 1. Introduction

Point defects are ubiquitous in crystalline solids and often play crucial roles in various fields of materials physics and chemistry[1,2]. Even when limited to non-metallic systems, which are our main target in this review, their applications span a wide range of materials, including semiconductors[3], biomaterials[4], ferroelectrics[5], ion conductors[6], (photo-)catalysts[7], and structural materials[8], to name a few. Despite their importance, it is generally difficult to fully investigate point defects experimentally because they are zero-dimensional in nature and exist on the atomic scale. Therefore, experimentalists usually infer defect characteristics using a combination of spectroscopic and analytical techniques, such as electron paramagnetic (or spin) resonance (EPR/ESR)[9], photoluminescence (PL)[10] and cathodoluminescence (CL)[11,12], deep-level transient spectroscopy (DLTS)[12], X-ray absorption fine structure (XAFS)[13], including extended X-ray absorption fine structure (EXAFS)[14], positron annihilation spectroscopy (PAS)[15], and secondary ion mass spectrometry (SIMS)[16].

To complement these experiments, first-principles calculations have become increasingly common [1]. Since the advent of hybrid functionals, which predict band gaps and localized defect states more accurately than the conventionally employed generalized gradient approximation (GGA) functionals such as PBE functional[17], the accuracy of first-principles calculations has improved dramatically. As a result, it is now possible to predict point defect properties even ahead of experiments. Indeed, we have employed the HSE hybrid functional to explain experimental findings in point-defect calculations [18–32]. However, defect calculations generally employ supercell models consisting of 60–300 atoms, and when combined with hybrid functionals, the computational cost becomes very high. In addition, it is necessary to consider a wide variety of defect types, along with their atomic configurations and charge states, the total number of which can easily exceed 100. Consequently, most studies focus on defects in only one or a few materials within a single paper (e.g., Ref. [33–46]).

Recently, high-throughput computational approaches have generated large volumes of materials data through repositories such as the Materials Project[47], AFLOW[48], and OQMD[49]. These datasets initially consisted mainly of previously reported crystal structures but have recently been expanded to include theoretically predicted stable structures[50]. Such data have been widely used to screen promising materials as well as to facilitate the construction of machine-learning (ML) models as training datasets. These ML models are sometimes employed to virtually screen promising materials from a vast number of candidates. Nevertheless, most screening efforts have so far been confined to bulk properties. This limitation arises primarily from the lack of defect-calculation datasets in major materials repositories. Another contributing factor is the relatively limited development of ML techniques for predicting point-defect properties.

Thus, screening superior materials from a point-defect perspective requires (i) well-established databases for point defects and (ii) further development of ML techniques to defect-related problems. The former issue remains largely unresolved; one exception is our oxygen-vacancy database, which comprises approximately 5,000 computational data points for neutral and positively charged oxygen vacancies in 927 oxides [51]. This database has been publicly available since 2021, and additional datasets have been released thereafter [52]. Such defect database can be directly used for screening superior materials [53,54]. Yet, most calculations are based on the GGA level, which may lead to erroneous predictions of localized states as mentioned above. In contrast, substantial progress has been made in recent years with respect to the latter issue; however, special care is still required when applying ML methods to defect-related problems because of challenges specific to point defects, such as charge states. Indeed, there is still no clear consensus on what has been achieved so far and what issues remain to be resolved in the future. This review provides a systematic overview of these points.

The ML applications in this field can be broadly categorized into two major classes. The first class focuses on the prediction of defect formation energies, while the second addresses the application of machine-learning potentials (MLPs) to point-defect problems, an area that has seen rapid progress in recent years. ML techniques can also be employed not only to predict the results of first-principles calculations, but also to analyze and interpret defect models or defects observed experimentally. For example, we identified characteristic features of oxygen vacancies in amorphous $SnO_2$ using ML [55]. Such studies are beyond the scope of this review.

In summary, this review overviews ML approaches for predicting point defect properties derived from first principles calculations, with an emphasis on defect formation energetics. In Section 2, we survey the major methodological families, including direct ML models that learn defect energetics from local structural descriptors and MLPs that enable efficient structural relaxation and finite temperature simulations. We also highlight representative model architectures, and the type of defect physics each approach can capture. Section 3 discusses remaining challenges and near-term opportunities, including dataset standardization, robust treatment of atypical relaxations, and consistent handling of charged defects through Fermi level alignment, finite size corrections, and long-range electrostatics. Section 4 closes with concluding remarks and a perspective on how these tools may be improved in the future.

## 2. Point-defect properties calculated from first principles

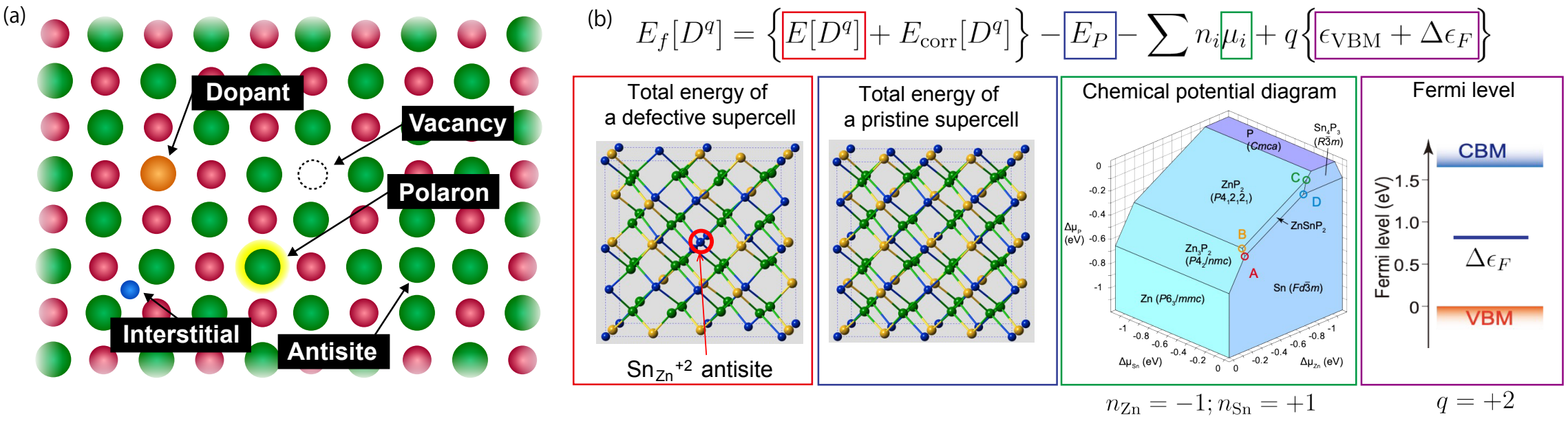

Fig. 1. (a) Schematic illustration of representative point defects in non-metallic materials, including vacancies, interstitials, antisites, dopants, and polarons. (b) Schematic depiction of the defect formation energy formalism, showing the total energy difference between defect and pristine supercells and the contributions from atomic chemical potentials, charge state and Fermi level, and finite size corrections (see text for details).

### 2.1. Types of point defects

Point defects are categorized into two primary groups, namely intrinsic and extrinsic defects, the latter of which are also referred to as impurities: whereas the former do not involve elements that are absent from the host structure, the latter do. The intrinsic defects include vacancies, antisites, and interstitials. In contrast, extrinsic defects include substitutional impurities and interstitials (see Fig. 1(a)). It should be noted that even when we refer to vacancies, not all vacancies correspond to simple vacancies. For example, upon vacancy formation, a neighboring host atom may move to a high-symmetry interstitial site, leaving behind two vacancies; this configuration is referred to as a split-type vacancy[51].

From another perspective, point defects can also be classified into two other classes: localized defects and defects that capture delocalized (hydrogenic) electrons or holes, which are also referred to as perturbed-host states [56]. In the latter case, for example, substitutional P in Si introduces an extra electron that is weakly bound to the P atom, with an energy level close to the conduction-band minimum (CBM). Such defects are therefore usually referred to as shallow defects. In supercell calculations, however, such hydrogenic (shallow or perturbed host) states are difficult to describe accurately, because the supercells typically employed are too small to capture their long-range character [57]. As a result, the calculated formation energies of such defects can suffer from significant finite-size errors. In practical defect calculations, because the binding energies of such states are not accurately evaluated, their transition levels are usually discussed only qualitatively. See Ref. [58–60] for more details. When applying ML methods to defect-related problems, special care must therefore be taken with such shallow defects.

### 2.2. Small polarons

In a solid, when an extra charge carrier is introduced, it interacts with the surrounding ions. When the electron–phonon interaction is sufficiently strong, the carrier can locally distort the lattice and form a deep electronic state, which is referred to as an electron (hole) polaron. In the case of a small polaron, the lattice distortion is typically confined to neighboring atomic sites. Such highly localized states can be modeled using relatively small supercells, similar to those employed for localized point defects [61–64]. Therefore, small polarons are included in this review as a special class of intrinsic defects. Note that when electron (hole) polarons exist without interacting with other defects, they are called self-trapped electrons (holes).

### 2.3. Defect formation energies

One of the most important quantities characterizing defects is their formation energy, because, except for nonequilibrium techniques such as irradiation or non-equilibrium growth, defect concentrations are governed by their formation energies. Electron exchange occurs between point defects, through changes in their charge states $q$, and the host valence and conduction bands, which act as reservoirs of holes and electrons, respectively. Because the charge-compensation mechanism generally cannot be ascertained *a priori*, it is common practice to first calculate the formation energies of all relevant defects as a function of the Fermi level, $\varepsilon_F$, and then determine the defect and carrier concentrations by enforcing charge neutrality.

The defect formation energy $E_\mathrm{f}$ is evaluated as

$$E_\mathrm{f}[D^q] = \{E[D^q] + E_\mathrm{corr}[D^q]\} - E_\mathrm{p} + \sum_i \Delta n_i \mu_i + q(\varepsilon_\mathrm{VBM} + \Delta\varepsilon_\mathrm{F}), \quad (1)$$

where $E[D^q]$ and $E_\mathrm{p}$ are the total energies of the supercell with a defect $D$ in $q$ and the corresponding pristine supercell, respectively (Fig. 1(b)). $\mu_i$ and $\Delta n_i$ are the chemical potential and the difference in the numbers of atom $i$ between the defect and perfect supercells, respectively. $\Delta\varepsilon_\mathrm{F}$ is the Fermi level to the valence band maximum (VBM), $\varepsilon_\mathrm{VBM}$. The chemical potentials depend on the growth conditions, and consequently the defect formation energies do as well. Naturally, Eq. (1) is constructed to conserve the numbers of constituent particles, such as electrons and atoms.

$E_\mathrm{corr}[D^q]$ corresponds to the correction term, which is required when we calculate the charged defects. For localized charged defects, the formation energy $E_\mathrm{f}[D^q]$ can be computed accurately provided that the finite-size correction $E_\mathrm{corr}[D^q]$ is properly included. Several schemes have been proposed to evaluate $E_\mathrm{corr}[D^q]$, and the Freysoldt–Neugebauer–Van de Walle (FNV) method [65] is now widely used as a standard approach. To extend its applicability, we have proposed an extended scheme that enables accurate correction energies for various defect types across diverse materials [66]. This scheme has been implemented in packages such as pydefect [51] and pycdt [67]. Note that for two-dimensional materials and surface slab models, specialized corrections are required because the dielectric response is spatially non-uniform. In addition, quantities such as defect eigenvalues, vertical transition energies, and configuration coordinate diagrams typically require different types of corrections [68–70]. We note that such corrections are applicable only to the localized defects, which are confined in the supercells adopted.

## 3. Machine learning of defect formation energies and related properties

### 3.1. Overview

When predicting defect formation energies using ML techniques, there are two main approaches (Fig. 2). In the first strategy ("direct ML model"), the defect formation energy is predicted *directly* for a given defect configuration. Here, the input typically consists of a pristine structure $X$ together with defect site index $i$ and the defect charge state $q$. The model is trained to map the local structural descriptors to the corresponding formation energy of a defect (Fig. 2(a)). The second strategy is to construct MLPs for defective and pristine structures (Fig. 2(b)). By approximating the DFT potential-energy surface, MLPs can predict energies and forces without explicitly solving the electronic structure. In this workflow, defect formation energies are obtained by relaxing the structure of the model with a defect using the MLP and then taking the energy difference between the relaxed defective and pristine systems. It should be noted, however, that many modern MLP frameworks are limited to charge-neutral simulations; consequently, formation energies can typically be evaluated only for neutral defects (light green line in right panel of Fig. 2). The accessible properties and the types of defects to which each strategy is applicable are compared in Table 1.

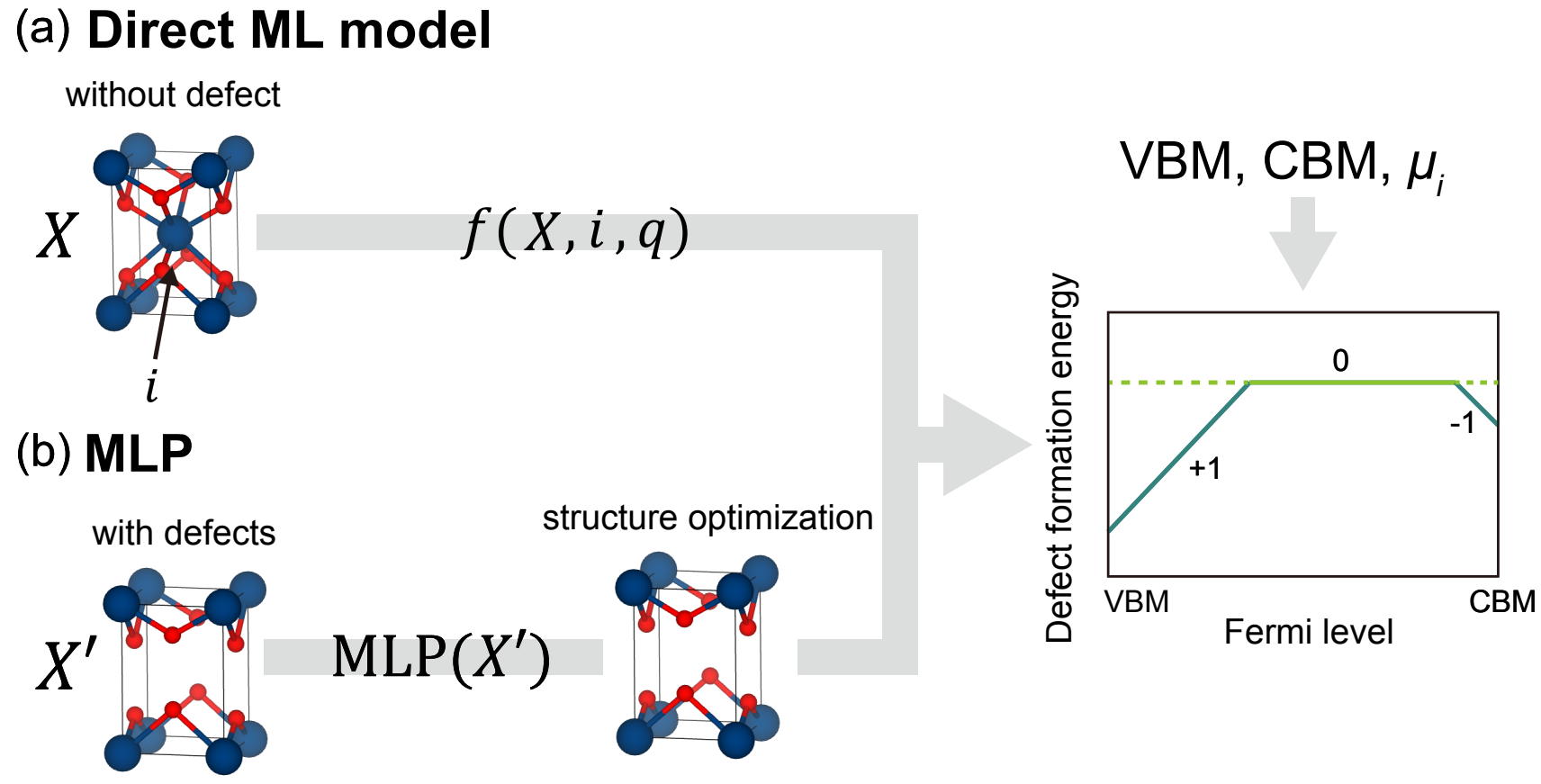

Fig. 2 Two ML strategies for predicting defect formation energies with an example of a vacancy created on the blue atom site: (a) direct ML models and (b) MLP based approaches. In the right panel, dark green lines are formation energies of charged defects and light green solid and dotted lines are those of neutral defects. $X$ and $X'$ are encoded structural information for the pristine and defective structures, respectively. $i$ denotes a site index in the crystal $X$ and $q$ denotes a charge of the defect. Additional information, including VBM, CBM, and $\mu_i$, respectively. In (a), $f(X, i, q)$ represents the regression model with input parameters: host structure $X$, defect site $i$, and charge state $q$.

Table. 1. Comparison of properties accessible by first-principles calculations, direct ML models, and MLPs. In the “Interstitials”, an asterisk (*) indicates that interstitial defects may be treated with direct ML models when the interstitial sites are predefined, such as in the zincblende structure.

| Method | Computational Cost | Formation energy | Atomic structure | Interstitials | Lattice dynamics | Electronic structure |
|---|---|---|---|---|---|---|
| First principles | High | ✓ | ✓ | ✓ | ✓ | ✓ |
| Direct ML models | Low | ✓ | × | ×* | × | × |
| MLPs | Medium | ✓ | ✓ | ✓ | ✓ | × |

In general, MLPs are computationally more expensive than direct ML models, but they remain far cheaper than first-principles calculations. The former typically enables prediction of formation energies only, whereas the latter can also provide relaxed atomic structures and lattice dynamics. In addition, the former cannot treat interstitials because their positions are generally not well-defined. However, there are exceptions when the target structural framework is fixed, such as in the zincblende structure, where well-defined tetrahedral and octahedral interstitial sites exist. In contrast, the latter can generally do so by considering several candidate interstitial sites followed by structure optimization. Note that neither approach can usually predict the electronic structure directly, such as the wavefunctions of localized electronic states or spin states. These quantities are, however, important for evaluating nonradiative carrier capture rates and the performance as quantum defects. Such electronic structures may be predicted using ML techniques designed to learn effective Hamiltonians, but further progress is needed for its application to a wide variety of defects.

In both approaches, a key difficulty in learning the formation energies of charged defects is their dependence on the Fermi level as shown in Eq. (1). We discuss how charge states should be treated in the following sections. Another important consideration is the scope of target materials. Some studies focus on a specific material to achieve high accuracy, whereas others aim to develop models applicable across a wide range of crystal structures. This generality–accuracy trade-off is unavoidable; however, general models are indispensable, especially for the virtual screening of promising materials. In the following sections, we individually review previous studies on direct ML and MLP approaches.

### 3. 2. Previous direct ML models for predicting defect formation energies

In direct ML models, defect properties are inferred from the local structures of the target sites. Table 2 summarizes previous studies on direct ML models, which are categorized according to the target materials, types of defects, defect properties, the exchange-correlation (XC) functionals used to generate the training data, whether charged defects are considered, types of descriptors, and the ML models employed. Here, we exclude the studies adopting linear regression (LR) models from the table, as they are usually not regarded as ML methods. Nevertheless, we emphasize that several LR-based studies have made significant contributions to the prediction of defect properties. As representative examples, we highlight the regression of neutral oxygen-vacancy formation energies in 45 oxides by Deml et al.[71] and the prediction of transition levels (TLs) of cation vacancies in 14 zincblende structures by Varley et al.[72]. As seen in Table 2, the ML applications to defects began around 2020. This progress was largely driven by the widespread availability of ML software packages. Below, we review previous studies from the perspective of each category.

- **Target materials / Defect types:** Several studies consider only one or a few compositions, and even when a larger number of materials is included, the structural frameworks are often limited, for example, to zincblende [73,74] or perovskite structures [75–77]. When a wide variety of compounds is considered, the target materials are mostly oxides, and the target defects are typically oxygen-vacancy formation energies[51,58,75,77–80]. This is primarily because oxygen vacancies constitute one of the most important classes of defects and their positions are relatively well defined. The first study to consider a wide range of structures was ours, published in 2021 [51]. The target defect types are diverse, including vacancies, antisites, and substitutional dopants. Interstitials are considered when the atomic frameworks are fixed as mentioned above. An interesting exception is the study by Birschitzky et al., in which small polarons in reduced $TiO_2$ and Nb-doped $SrTiO_3$ were investigated[81].
- **Target properties / Charged defects:** Only defect formation energies and/or TLs have been investigated so far. Regarding TLs, Frey et al. identified whether defects are deep or shallow [82], which is a classification problem. Because the prediction of TLs requires consideration of charged defects, two main approaches have been explored: one is to directly predict TLs, and the other is to predict the formation energies of charged defects and subsequently derive TLs from these energies. When TLs are predicted directly relative to the VBM, the information on VBM position is also included in the TLs, which is generally more challenging than predicting defect formation energies alone. In the latter approach, the dependence on the Fermi level must be properly treated, which becomes particularly challenging when dealing with a wide variety of defects; this issue is discussed in detail later. Moreover, charged-defect calculations in DFT require finite-size corrections to the defect formation energies. Consequently, most studies focus on neutral defects.

- **XC functionals:** Most studies adopt GGA functionals (PBE [17] or PBEsol [83]), and some additionally apply Hubbard $U$ corrections[84] to mitigate delocalization errors. Wexler *et al.* used SCAN+$U$ for oxygen vacancies in perovskite oxides[75], and several studies [85–88] employed the HSE hybrid functional[89]. Note that, in some defects, only hybrid functionals can correctly describe localized defect states, including polarons [90], although their computational cost is significantly higher than that of GGA or meta-GGA functionals. Exceptions include oxygen vacancies that can usually be described reasonably well even with GGA functionals[51]. This is one of the reasons why high-throughput point-defect calculations mostly focus on oxygen vacancies.
- **Descriptors / ML models**: When constructing ML models, descriptors are first extracted from the local structures around the target defects. In this review, we classify descriptors into two types. The first class consists of physical descriptors, which are constructed from elemental (*e.g.*, electronegativity and ionic radii), structural (*e.g.*, Bader volume and coordination number), and host-related information (*e.g.*, band gaps and dielectric constants). These descriptors are often handcrafted based on prior expert knowledge. For example, oxygen-vacancy formation energies have been shown to correlate with oxide formation energies [71] and band gaps [91,92]. Such descriptors can be combined with a wide variety of ML models, including RFR [93], kernel ridge regression (KRR) [94], gradient-boosting regression (GBR) [95], support-vector regression (SVR) [96], and least absolute shrinkage and selection operator (LASSO) [97]. An important advantage of the physical descriptors is that they enable analysis of the key factors governing defect formation energies, for instance by examining feature importances in RFR.

  The second class comprises graph-based representations, which are combined with neural-network models. They have been widely developed and adopted in recent years. Examples include the crystal graph convolutional neural networks (CGCNN)[98], materials graph network (MEGNet)[99], and atomistic line graph neural network (ALIGNN)[100]. These representations have increasingly replaced traditional physical descriptors because they can be constructed without prior domain knowledge and often achieve higher predictive accuracy.

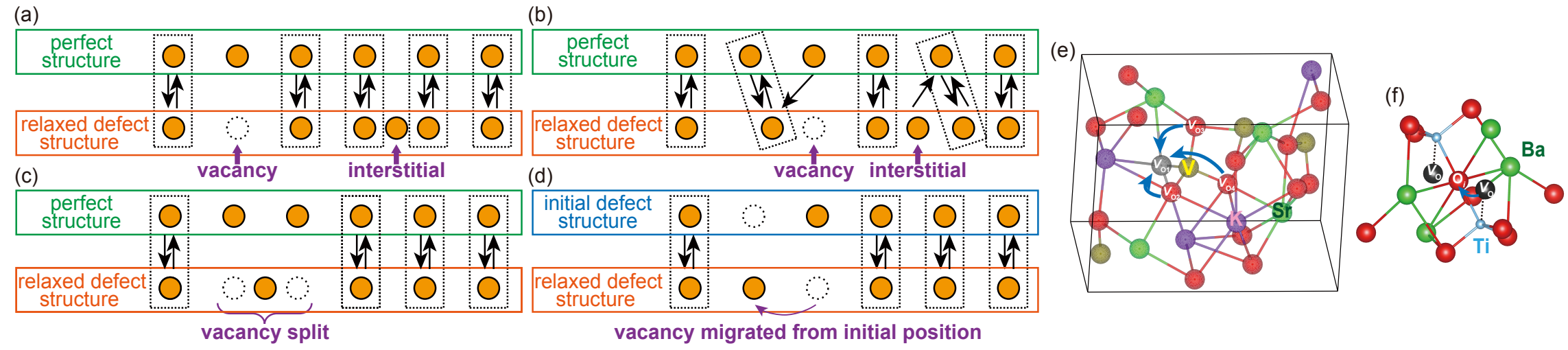

FIG 3 (a–d) Schematic illustration of the atom pairing technique. Orange spheres denote atoms of the same element in the supercell. A one-to-one atom correspondence is established (a–c) between the perfect and relaxed defect structures and (d) between the initial and relaxed defect structures. Paired atoms are indicated by bidirectional arrows and enclosed by dashed boxes. (e) Crystal structure of $KSrVO_4$, highlighting the four inequivalent O sites surrounding a V ion. Vacancy migration from three sites is indicated by blue arrows. (h) Local structure of a split type oxygen vacancy in $Ba_2TiO_4$. When the left vacancy is introduced, the neighboring right O ion migrates to a high symmetry site. Reprinted from Ref. [51] licensed under CC BY 4.0.

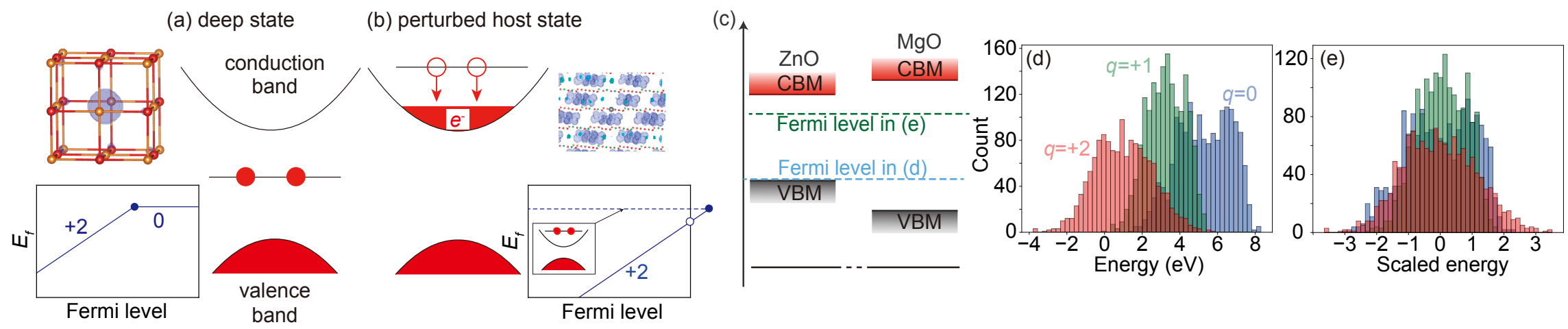

FIG 4 Schematic illustration of defects with localized occupied states (a) within the band gap and (b) above the conduction band minimum (CBM). In (b), the electrons relax to the CBM, resulting in delocalized perturbed host states (PHS). The formation energy of a defect with PHS is indicated by a dashed line. Donor transition levels are schematically shown in the formation energy diagrams as a function of the Fermi level. Transition levels involving deep localized states can be computed reliably from first principles (solid circles). In contrast, shallow levels associated with PHS are shown only qualitatively (open circle), because they cannot be computed using realistically sized supercells (see text for details). Squared wave functions for these states in neutral oxygen vacancies in MgO and $BaTiO_3$ are also shown as examples. (c) Schematic illustration of core potential alignment using ZnO and MgO as examples. Even after aligning eigenvalues based on oxygen core potentials, the absolute Fermi level remains arbitrary. Here, we propose

determining the Fermi level by maximizing the overlap between the distributions of oxygen vacancy formation energies. (d) Distributions of oxygen vacancy formation energies for $q$ = 0, +1, and +2 with $\epsilon_F$ aligned to the VBM of ZnO via core potentials. Data for vacancies exhibiting PHS have been removed. (e) Normalized distributions after optimizing $\epsilon_F$ to minimize the mean differences between charge states. Reprinted from Ref. [58], licensed under CC BY 4.0.

These previous studies have contributed significantly to the development of ML approaches for defect formation energies and TLs. However, the quality of the training dataset has not always been carefully considered, despite its pivotal role in constructing high-accuracy ML models. When applying ML to defect-related problems, particular care must be taken with respect to the following points.

1. The first concerns atomic structures. When direct ML models are applied, they implicitly assume typical substitutional defects and vacancies, meaning that neighboring atoms undergo only moderate displacements and that defect positions are well defined. However, split-type vacancies in Si and DX-center–type configurations in zincblende and chalcopyrite structures, which involve large displacements of neighboring atoms and deviate significantly from typical substitutional defects and vacancies, can exhibit the lowest energies. Whether such configurations should be included in the training dataset is generally a nontrivial question. If these datasets are excluded, the predicted formation energies should be regarded as upper limits rather than definitive values. Another important difficulty arises from defect migration, where defects relax to different lattice sites during structural optimization. Such cases blur the definition of the target defect and should therefore be carefully identified and, in many cases, excluded from the training dataset. The identification of these unusual structural defects can be achieved by using the atom mapping technique which is depicted in Fig. 3 and described in details in Ref. [51].
2. In addition, we classify defects according to whether they host deep localized states or perturbed host states (PHS). As discussed in detail in Refs. [51,56,58,59], PHS correspond to shallow, hydrogenic states whose wavefunctions extend over length scales far exceeding typical supercell sizes. As a result, their formation energies and transition levels cannot be evaluated accurately within standard supercell calculations, making them unsuitable as reliable training data for ML models (see Figs. 4(a) and (b)). Therefore, defects with PHS should be excluded from the dataset.

Indeed, we have demonstrated that excluding such defects improves the prediction accuracy of oxygen-vacancy formation energies. In addition, because the constructed models can predict the transition levels, they can consequently determine where PHS are generated [58].

3. The determination of the Fermi level becomes another critical issue for charged defects, since their formation energies depend linearly on the Fermi level. In particular, when targeting a wide range of materials, a consistent treatment of Fermi-level alignment is essential for constructing transferable ML models. In our previous study on the prediction of oxygen-vacancy formation energies, we proposed aligning the Fermi levels using the oxygen core potential as a common reference as shown in Fig. 4(c). This approach is physically motivated, as it removes spurious electrostatic potential contributions. The remaining arbitrariness in the Fermi-level position is then determined so as to maximize the overlap of formation-energy distributions across different charge states (Figs. 4(d) and (e)). This procedure facilitates the standardization of target values, which is known to improve weight updates during ML training.

Table 2. Summary of the studies on direct ML methods. In the "Target properties" column, FE and TL represent defect formation energies and transition levels, respectively. In the "XC functional" column, PBE, HSE, and SCAN denote the Perdew–Burke–Ernzerhof, Heyd–Scuseria–Ernzerhof, and "Strongly Constrained and Appropriately Normed" functionals, respectively, and "+U" indicates the inclusion of Hubbard U corrections. In the "Descriptors" column, PD, CGR, SD, SF, and PHF represent physical descriptors, crystal graph representations, structural descriptors, symmetry functions, and persistent homology features, respectively. In the "ML models" column, the abbreviations denote the following methods: GPR (Gaussian process regression), NN (neural network), RF/RFR (random forest/random forest regression), KRR (kernel ridge regression), LASSO (least absolute shrinkage and selection operator), SVM (support vector machine), SVR (support vector regression), LR (linear regression), GBR (gradient boosting regression), KNR (k nearest neighbors regression), CGCNN (crystal graph convolutional neural network), MEGNet (materials graph network), ALIGNN (atomistic line graph neural network), and GNN (graph neural network). "etc." indicates additional models not explicitly listed.

| First authors and Ref | Publish year | Host materials (Number of compositions) | Type of defects | Target Properties | XC functional | Charged Defects | Descriptors | ML models |
|---|---|---|---|---|---|---|---|---|
| Batra[101] | 2019 | $HfO_2$ (1) | Substitutional dopants | FE | PBE | No | PD | GPR, etc |
| Cheng[102] | 2020 | GeTe (1) | Ge vacancies | FE | PBE | No | SF | NN |
| Sharma[103] | 2020 | $BaTiO_3$ and $LaMnO_3$ (2) | Substitutional dopants | FE | PBE | No | PD | RFR |
| Frey[104] | 2020 | 2D materials (158) | Vacancies, divacancies, | FE / TL (deep or shallow) | PBE | No | PD | RFR |

| | | | antisites, dopants | | | | | |
|---|---|---|---|---|---|---|---|---|
| Mannodi-Kanakkithodi[85] | 2020 | Cd*X* (*X*=S, Se, Te) (3) | Dopants | FE / TL | PBE/HSE | TL | PD | RFR, KRR, LASSO |
| Wan[78] | 2021 | Metal oxides (1750) | Oxygen vacancies | FE | -- | No | PD | SVM, GPR, NN, etc |
| Wexler[75] | 2021 | Oxide perovskites (233) | Oxygen vacancies | FE | SCAN+U | No | PD | LR |
| Kumagai[51] | 2021 | Metal oxides (937) | Oxygen vacancies | FE | PBESol+U | Yes | PD | RFR |
| Mannodi-Kanakkithodi[73] | 2022 | Zinc blende semiconductors (34) | Vacancies, substitutions and interstitials | FE / TL | PBE | TL | PD | LR, GPR, RFR, NN, KRR |
| Polak[86] | 2022 | Zinc blende semiconductors (14) | Dopants | TL | LDA/PBE/HSE06 | TL | PD | GBR |
| Birschitzky[81] | 2022 | $TiO_2$ and Nb-doped $SrTiO_3$ surfaces | Small polarons and charged point defects | FE | PBE+U | No | SD | KRR |

| Wu[76] | 2023 | Halide perovskites (38) | Vacancies, interstitials, antisites | TL | --- | TL | Various PD | GBR, SVR, KNR |
|---|---|---|---|---|---|---|---|---|
| Sharma[87] | 2023 | $BaZrS_3$ (1) | Cationic dopants | FE | PBE/HSE06 | No | PD | RFR and CGCNN |
| Baldassarri[79] | 2023 | Metal oxides (1157) | Oxygen vacancies | FE | PBE | No | PD / CGR | RFR, SVR, KRR, LR |
| Witman[105] | 2023 | Metal oxides (~200) | Oxygen vacancies | FE | PBE+U | No | CGR | CGCNN |
| Rahman[74] | 2024 | Zinc blende semiconductors (34) | All types | FE / TL | PBE | TL | CGR | CGCNN, MEGNet, ALIGNN |
| Park[80] | 2024 | Metal oxides (~1000) | Oxygen vacancies | FE | PBE+U | No | CGR | MEGNet, CGCNN |
| Kiyohara[58] | 2025 | Metal oxides (932) | Oxygen vacancies | FE | -- | Yes | CGR | CGCNN |
| Fang[77] | 2025 | Oxide perovskites (~1100) | Vacancies, substitutions | FE | PBE+U-D3 | No | CGR | Various GNN models |
| Khamdang[88] | 2025 | $CsSnI_3$ (1) | native defects and dopants | FE | HS06 | Yes | CGR w/ PHF | RF, GPR, KRR, LR |

### 3.3 Previous MLPs for predicting defect energetics

Direct ML methods enable fast prediction of defect formation energies but typically provide no information about atomic structures, essential for understanding defect properties. For example, defect concentrations depend on configurational degeneracy, which is related to defect site symmetry. In addition, phonon properties near point defects provide information beyond formation energies, such as defect free energies and ionic diffusion. Obtaining these quantities is often computationally expensive because they require phonon calculations and MD simulations, respectively, which are particularly costly at the first-principles level. MLPs are among the most promising approaches for addressing these limitations. By approximating the potential energy surface obtained from first-principles calculations, MLPs can predict energies and forces without explicitly solving the electronic structures. As a result, they can achieve near *ab initio* molecular dynamics (AIMD) level accuracy at a much lower computational cost, enabling not only structure optimization but phonon calculations and long-time MD simulations.

Let us start with the history of the conventional MLPs whose targets are mainly bulk solids. Various types of MLPs have been developed so far, differing mainly in how atomic environments are represented and what regression methods are adopted. Early and widely used approaches employ physically motivated basis functions or descriptors, such as those introduced in the Behler–Parrinello framework, combined with general-purpose neural networks[106]. In contrast, kernel-based methods, exemplified by Gaussian Approximation Potentials[107] (GAP), model similarities between atomic environments using kernel functions and often provide high accuracy with strong data efficiency. More recently, graph neural network (GNN)–based approaches have been proposed[108–111], in which crystal structures are treated as graphs and descriptors are learned directly from data with fewer human-designed bias. These models have been further extended by explicitly incorporating equivariance with respect to rotations, enabling strict enforcement of geometric symmetries and improving both physical consistency and representational power[112–114]. For ionic materials, long-range (LR) Coulomb interactions play a crucial role in accurately describing atomic interactions. To account for these effects, recent MLP frameworks explicitly incorporate electrostatic interactions[115–119], typically through Ewald methods. By coupling short-range MLPs with physically motivated LR Coulomb interactions, the transferability and accuracy of MLPs for ionic systems are enhanced. Furthermore, beyond the prediction of energies and forces, MLPs can be naturally extended to output other energy-derivative properties, which may be learned simultaneously. By sharing a common representation of atomic environments across multiple targets, such models can capture correlations between structure and different physical properties in a unified manner. In particular, the simultaneous learning of polarization and Born effective charges alongside energies and forces provides a consistent framework for modeling external field-responsive and dielectric properties at the atomic scale[120,121].

MLPs have also begun to be applied to the study of point defects and defect-driven phenomena in solids. Their use now spans a wide range of research areas, including ionic conductors[122–135], semiconductors[136–147], thermal conductors[148–158], and related systems. In the following, we review previous studies by category and summarize them in Table 3.

- **Types of point defects and charge states**: Previous studies have investigated a wide variety of point defects; however, comparatively less attention has been paid to charge neutrality and defect charge states. Treating charged defects generally requires either (i) consideration of multiple defects to ensure overall charge neutrality, or (ii) introduction of a uniform compensating background charge to enforce neutrality of a supercell. Niu *et al.* [131] provide an example of (i): they simulated proton conductivity while maintaining charge neutrality via Y dopants on Zr sites. Birschitzky *et al.* [159] present a representative case of (ii), in which polaron simulations were performed by introducing an extra electron or hole, with a compensating background charge.
- **Target properties**: Many studies have focused on dynamical properties such as activation energies and thermal conductivity. For ionic conductors, MLPs enable simulations of point-defect diffusion on spatial and temporal scales far beyond those accessible to AIMD. Likewise, in the context of thermal transport, MLPs permit simulations of phonon scattering by point defects at a computational cost low enough to make higher-order interactions tractable. For semiconductor applications, the defect formation energy remains a key descriptor. In this regard, Mosquera-Lois *et al.* [137] proposed a framework that accounts for entropic contributions to defect free energies at finite temperatures.
- **Class of MLP**: The MLPs used in the defect studies span a wide range of model classes, from descriptor-based MLPs, including kernel-based approaches (GAP[107]) and linear or polynomial models built on fixed local descriptors (MTP[160]), to conventional neural-network potentials that also rely on predefined local descriptors (Behler–Parrinello MLP[106]), and more recent message-passing and equivariant architectures (SchNet [109], NequIP [113], Allegro [114], MACE [112], M3GNet [110], CHGNet [111]). While these approaches differ in representation and learning algorithms, most share a common structural assumption: the energy is expressed as a sum of local atomic contributions within a finite cutoff. This locality enables efficient, large-scale simulations of defect dynamics, but it also means that long-range electrostatics are often not treated explicitly unless it is added as a separate term or handled through an auxiliary model. In addition, several studies emphasize that the data workflow and training strategy, particularly active-learning pipelines such as DP-GEN[161], can be as critical as the architecture choice for achieving transferability across diverse defect environments.

Although many studies have reported the application of MLPs to point defects, three issues still require special attention.

- As discussed in the Section 3.2, the treatment of charged defects remains a challenging issue. Two commonly used strategies mentioned above entail limitations. In approach (i), charge compensation inevitably introduces defect–defect interactions, making it difficult to separate the contribution of the target defect from that of the compensating species. To disentangle them, approach (ii) is straightforward. However, in such charged supercell calculations, the total energy depends on the chosen reference level of the electrostatic potential and the Fermi level. As a result, the total energy is not uniquely defined and cannot be used directly as training data. In practice, previous studies have not applied any special treatment to this issue, because they used supercells of fixed size and assumed that such ambiguities are absorbed into a constant energy contribution. In such cases, however, it would be difficult to predict defect formation energies across different supercell models, because the reference energy levels may change.
- Long-range Coulomb interactions play a crucial role, particularly for charged defects. Therefore, an appropriate treatment of dielectric screening, which reduces these interactions, is also essential.
- In some vacancies, electride-like states, in which excess electrons localize around the vacancy sites, may be stabilized. These states can behave similarly to ions, and therefore the descriptors need to include information on the positions of the electrides.

Table 3. Summary of MLP applications to point defects. The "First authors and Ref" column lists the first author of each study together with the corresponding reference. The "Year" column indicates the publication year. The "Host Materials (count)" column summarizes the host materials investigated in each study, with the number of distinct compositions given in parentheses where applicable. The "Type of Defects" column specifies the kinds of defects considered, vacancies (V), interstitials (I), antisites (A), or extrinsic dopants (E). The "Target properties" column lists the main target properties predicted in each work. The "XC functional" column indicates the exchange-correlation functional used in the underlying first-principles calculations. Abbreviations consistent with those used in Table 2 are adopted here, whereas functionals not listed in Table 2 are accompanied by the corresponding references. The "Charged Defects Considered" column shows whether charged defect were included in the study. The "ML Models" column summarizes the machine-learning model architectures adopted in each work. Models that coincide with those introduced in Sec. 3.3 are denoted using the same terminology, whereas other or study-specific architectures are accompanied by references in parentheses, where the term in parentheses indicates the underlying base model. Here, NN denotes a conventional multilayer neural network. The "LR Coulomb interactions Considered" column indicates whether LR Coulomb interactions were explicitly taken into account to the MLP.

| First authors and Ref. | Year | Host Materials (count) | Type of Defects | Target properties | XC functional | Charged Defects Considered | ML Models | LR Coulomb interactions Considered |
|---|---|---|---|---|---|---|---|---|
| Liang [162] | 2017 | $Ta_2O_5$ | E | Diffusivity | PW91[163,164] | No | NN | No |
| Bartók [165] | 2018 | Si | V | Diffusivity /Formation energy | PW91[166] | No | GAP | No |
| Lacivita [123] | 2018 | LiPON | V | Diffusivity | PBE | No | NN | No |

| Miwa [124] | 2018 | $Li_7La_3Zr_2O_{12}$ | E | Diffusivity | Wu and Cohen-GGA[167] | No | NN | Yes |
|---|---|---|---|---|---|---|---|---|
| Babaei [152] | 2019 | Si | V | Phonon | PBE | No | GAP | No |
| Deng [125] | 2019 | $\alpha$-$Li_3N$ | V, I | Diffusivity /Phonon/Formation energy | PBE | No | eSNAP [125] (Spectral Neighbor Analysis Potential) | Yes |
| Wang [122] | 2020 | Li-cathode coating materials (181) | V | Diffusivity | PBE | Yes | MTP | No |
| Rao [126] | 2020 | $Li_{10}GeP_2S_{12}$ / $Li_{10}SiP_2S_{12}$ | E | Diffusivity | PBE | No | NN | No |
| Rowe [168] | 2020 | C | V | Formation energy | optB88-vdW[169] | No | GAP | No |
| Houchins [127] | 2020 | $LiNiO_2$/$LiCoO_2$/$LiMnO_2$ | V | Free energy/Open circuit voltage | BEEF-vdW[170] | No | Behler-Parrinello MLP | No |
| sLi [128] | 2020 | $Na_3OBr$ | V | Diffusivity | PBEsol | No | DP-GEN | No |
| Qi [129] | 2021 | LLTO/$Li_3YCl_6$/$Li_7P_3S_{11}$ | V | Diffusivity /Phonon | optB88-vdW [169] | No | MTP | Yes |

| Shimizu [136] | 2022 | GaN | V | Phonon/Formation energy | PBE | Yes | Behler-Parrinello MLP | No |
|---|---|---|---|---|---|---|---|---|
| Dai [130] | 2022 | $Li_xLa_3Zr_{x-5}Ta_{7-x}$ $xO_{12}$ | V | Diffusivity | PBEsol | No | NN | No |
| Niu [131] | 2022 | $BaZrO_3$ | E | Diffusivity | PBE | No | DeePMD | No |
| Zhang [153] | 2022 | $Bi_2Te_3$ | V/A | Phonon/Formation energy | PBE-D3[171] | No | DeePMD | No |
| Ding [139] | 2022 | Si/GaN | E | Diffusivity /Formation energy | PBE | No | DeePMD | No |
| Lama [132] | 2023 | $LiFePO_4$ | E | Diffusivity | PBE+U | No | VASP-FF[172,173] | No |
| Zhang [154] | 2023 | $Sb_2Te_3$ | E | Phonon | PBE-D3 [171] | No | DeePMD | No |
| Ghim [155] | 2023 | β-GeTe | V/A | Phonon | - | No | NequIP | No |
| Bocus [174] | 2023 | zeolite | E | Free energy/Nuclear quantum effects | revPBE-D3 [171,175] | No | SchNet | No |
| Pols [140] | 2023 | $CsPbI_3/CsPbBr_3$ | V | Diffusivity | GGA | No | Sparse Gaussian | No |

| | | | | | | | Process[176,177] | |
|---|---|---|---|---|---|---|---|---|
| Han [178] | 2023 | ZnO-surface | V | Diffusivity /Free energy | PBE | No | EANNP [179] | No |
| Wu [180] | 2023 | $TiO_2$ | V | Diffusivity | SCAN | No | DeePMD | No |
| Jiang [181] | 2024 | $ThO_2$ | V/I | Formation energy | LDA | Yes | NN | No |
| Klarbring [133] | 2024 | $Na_3SbS_4$ | V/E | Diffusivity | $r^2$SCAN[182] | No | Allegro | No |
| Lu [134] | 2024 | $LiNiO_2$ | V/E | Formation energy | PBE | No | M3GNet | No |
| Liu [143] | 2024 | $CsPbI_3$ | V/I/A | Capture rate | PBE | No | DeePMD | No |
| Zhang [156] | 2024 | SnSe | V | Phonon | PBE-D3 [171] | No | DeePMD | No |
| Zhang [157] | 2024 | BAs | V | Phonon | PBE | No | DeePMD | No |
| Mosquera-Lois [144] | 2024 | semiconductors (50) | V | Structure optimization | HSE06 | No | M3GNet, CHGNet, MACE | No |
| Chen [183] | 2024 | $HfO_2$ | V/A/I | Radiation damage | PBE | No | NN | No |
| Liu [184] | 2024 | 3C-SiC | V | Phonon/Radiation damage | PBE | No | DP-ZBL [185](NN) | No |
| Hudson [186] | 2024 | $BaZrO_3$ | V/E | Phonon/Free energy | PBE | No | DeePMD | No |
| Mosquera-Lois [137] | 2025 | CdTe | V/I | Free energy | PBEsol | Yes | MACE | No |
| Dou [148] | 2025 | AlN | V/I/A | Phonon | PBE | Yes | Behler-Parrinello MLP | No |
| Sharma [149] | 2025 | 2D semiconductors (10) | E | Phonon | - | Yes | M3GNet, MACE, SevenNet, ORB[187](EGNN), MatterSim[188] | No |
| Zhou [150] | 2025 | GaN/ZnO | E | Phonon | PBE | Yes | NequIP | No |
| Liang [189] | 2025 | FeO | V | Formation energy | PBE+U | Yes | DeePMD | No |
| Birschitzky [159] | 2025 | MgO/$TiO_2$ | E/hole/electron | Diffusivity | PBE+U | Yes | LEOPOLD [159] (NequIP) | No |

| Yang [151] | 2025 | GaN | V/I/A | Phonon/Formation energy | LDA[190] | Yes | DP-GEN | No |
| --- | --- | --- | --- | --- | --- | --- | --- | --- |
| Tyagi [138] | 2025 | $CsPbI_3$ | V/I | Diffusivity /Formation energy | PBE-D3-BJ[191] | Yes | GAP | No |
| Wang [192] | 2025 | $CaSiO_3$ | V | Diffusivity /Free energy | PBE | No | DeePMD | No |
| Yokoi [158] | 2025 | PbTe | V/A/I | Phonon | PBEsol | No | Behler-Parrinello MLP | No |
| Wisesa [193] | 2025 | MgO/AgO/CuO | V | Formation energy | PBE | No | DeePMD, MTP | No |
| Zhou [135] | 2025 | $Ba_7Nb_4MoO_{20}$ /$Sr_3V_2O_8$ | V/E | Diffusivity | PBE | No | MTP | No |
| Arber [146] | 2025 | γ-$CsPbI_3$ | V/E | Diffusivity | PBEsol | No | MACE | No |
| Yang [194] | 2025 | 2D semiconductors (6) | V/E | Structure optimization | PBE | No | DefiNet [194] (EGNN) | No |
| Yan [147] | 2026 | C(diamond) | E | Formation energy | PBE | No | DeePMD | No |
| Wang[195] | 2026 | $Sb_2Se_3$ | V | Energy/Force | PBE, HSE06 | Yes | MACE | No |

## 4 Summary and outlook

ML is rapidly reshaping how point defects in non-metallic materials are studied, especially for quantities traditionally obtained from first-principles calculations, such as defect formation energies and related thermodynamic parameters. Across the literature, two complementary directions stand out. Direct ML models aim to predict defect energetics from representations of local environments and are well suited for rapid screening, whereas MLPs approximate the potential energy surface and enable large-scale structural relaxation and finite-temperature simulations. Their strengths are distinct: direct ML is fast but has limited access to dynamics and complex relaxations, whereas MLPs can address kinetics and temperature effects but require more extensive data and careful validation. A likely use case is a tiered workflow, in which direct ML is used for fast screening and MLP-based simulations are used to connect energetics to temperature-dependent thermodynamics and transport. Such defect-aware screening may enable high-throughput materials discovery based on defect properties.

A key limitation of ML-based prediction is not only model capacity, but also the reliability and definition of the training labels. Defect datasets may include atypical relaxations and structural transformations, such as split configurations, DX-like reconstructions, or relaxations that effectively change the defect site. They may also include shallow defects with perturbed-host states, which are not well described by standard supercell methodologies. Such unintended defect data can reduce accuracy and obscure what the model actually learns, making dataset curation and defect-state classification as important as the learning algorithm itself.

The most consequential frontier is charged-defect physics. Formation energies depend on Fermi-level alignment and finite-size corrections, and inconsistent reference choices can make learned targets incompatible across materials or studies. Progress will benefit from community-standardized protocols for alignment and corrections, as well as from ML targets designed to be robust under these protocols, such as transition levels or charge-dependent energetics defined within a unified reference scheme. In addition, a general-purpose MLP should yield the same defect formation energy regardless of supercell size; this size independence is one criterion for a physically correct MLP for defects. In parallel, long-range electrostatics and dielectric screening require more explicit treatment, because the locality assumptions built into many models can fail in the dilute limit or when Coulomb interactions dominate.

**Acknowledgement**
This work has been supported by JST FOREST Program (JPMJFR235S), KAKENHI (25K01486) and the Asahi Glass Foundation.